\documentclass[11pt]{article}
\usepackage{UF_FRED_paper_style}

\title{Modeling Fission Gas Release at the Mesoscale using Multiscale DenseNet Regression with Attention Mechanism and Inception Blocks}

\author{Peter Toma\\% Name author
    \href{mailto:ptoma@ufl.edu}{\texttt{ptoma@ufl.edu}} \\ %% Email author 1 
\and Md Ali Muntaha\\% Name author
    \href{mailto:md.muntaha@ufl.edu}{\texttt{md.muntaha@ufl.edu}} \\ %% Email author 2
\and Joel B. Harley\\% Name author
    \href{mailto:joel.harley@ufl.edu}{\texttt{joel.harley@ufl.edu}}%% Email author 3
    \\
\and Michael R. Tonks\thanks{Corresponding Author: Michael R. Tonks, michael.tonks@ufl.edu}\\% Name author
    \href{mailto:michael.tonks@ufl.edu}{\texttt{michael.tonks@ufl.edu}} \\ %% Email author 4
    }
    
\date{\today}

\begin{document}
\onecolumn
% %%%%%%%%%%%%%%%%%%%%%%%%%%%%%%%%%%%%%%%%%%%%%%%%%%%%%%%%%%
% %%%%%%%%%%%%%%%%%%%%%%%%%%%%%%%%%%%%%%%%%%%%%%%%%%%%%%%%%%
% ABSTRACT
% %%%%%%%%%%%%%%%%%%%%%%%%%%%%%%%%%%%%%%%%%%%%%%%%%%%%%%%%%%
% %%%%%%%%%%%%%%%%%%%%%%%%%%%%%%%%%%%%%%%%%%%%%%%%%%%%%%%%%%
{\setstretch{.8}
\maketitle
% %%%%%%%%%%%%%%%%%%
\begin{abstract}
% CONTENT OF ABS HERE--------------------------------------

Mesoscale simulations of fission gas release (FGR) in nuclear fuel provide a powerful tool for understanding how microstructure evolution impacts FGR, but they are computationally intensive. In this study, we present an alternate, data-driven approach, using deep learning to predict instantaneous FGR flux from 2D nuclear fuel microstructure images. Four convolutional neural network (CNN) architectures with multiscale regression are trained and evaluated on simulated FGR data generated using a hybrid phase field/cluster dynamics model. All four networks show high predictive power, with $R^{2}$ values above 98\%. The best performing network combines a Convolutional Block Attention Module (CBAM) and InceptionNet mechanisms to provide superior accuracy (mean absolute percentage error of 4.4\%), training stability, and robustness on very low instantaneous FGR flux values.\\

% END CONTENT ABS------------------------------------------
%% <-- Keywords HERE!
\noindent

\end{abstract}
}

% %%%%%%%%%%%%%%%%%%%%%%%%%%%%%%%%%%%%%%%%%%%%%%%%%%%%%%%%%%
% %%%%%%%%%%%%%%%%%%%%%%%%%%%%%%%%%%%%%%%%%%%%%%%%%%%%%%%%%%
% BODY OF THE DOCUMENT
% %%%%%%%%%%%%%%%%%%%%%%%%%%%%%%%%%%%%%%%%%%%%%%%%%%%%%%%%%%
% %%%%%%%%%%%%%%%%%%%%%%%%%%%%%%%%%%%%%%%%%%%%%%%%%%%%%%%%%%

% --------------------
\section{Introduction}
% --------------------

% \subsection{Fission Gas Release}

While only comprising 4.3\% of global energy production in 2020 \cite{BP2021}, nuclear power has gathered renewed interest in recent years for offering significant quantities of reliable, low carbon footprint electricity. The first civilian nuclear reactors for electricity generation debuted in the 1950s, but there remains significant unsolved challenges regarding their long-term operation. One such issue is the problem of fission gas release (FGR), most known in the polycrystalline UO$_{2}$ fuel pellets used by modern commercial light water reactors (LWRs) \cite{Tonks2018}. FGR, occuring when a fission reaction produces noble gases such as xenon and krypton as waste products, is an inevitable byproduct of the operation of a LWR. 

Fission products do not dissolve into the nuclear fuel microstructure but instead form gas bubbles inside the microstructure. These bubbles reduce the thermal conductivity of the fuel \cite{Tonks2018}, reducing the efficiency with which heat can be converted to electricity. Intergranular fission gas bubbles grow and interconnect, eventually providing a path for FGR when gas escapes the fuel into the cladding. This decreases heat transport through the gap between the fuel and cladding, and increases the cladding pressure. The inability to remove heat from the fuel accelerates the degradation of its microstructure by FGR even further, thus establishing a negative feedback loop. Currently, one third of the nuclear fuel rods in standard LWR reactors must be replaced every 12-24 months \cite{EIA2022} - a severe limitation given the difficulty of shutting down and restarting a nuclear reactor. Thus, understanding the process of FGR is critical to increasing the efficiency and safety of LWR fuel.

The traditional method of estimating FGR from nuclear fuel microstructures is by reduced order models that approximate the physics underlying FGR \cite{Tonks2018,Rest2019}. More recently, mesoscale models have been developed that spatially resolve the fission gas bubble evolution and give a more accurate description of the fission gas behavior \cite{Tonks2018}, and the phase-field method has emerged as one of the most popular approaches for these mesoscale simulations \cite{hu2009phase,Millet_2012_a,Larry_2019,PRUDIL2022153777,Tonks2022, muntaha2023}. One limitation of the phase field method is that it is not computationally feasible to model larger intergranular bubbles and small intragranular bubbles in the same simulation. This has been overcome by a recent hybrid model that couples the phase field method of intergranular bubbles with a spatially resolved cluster dynamics model of intragranular fission gas \cite{Tonks2022}. 

Presently, the hybrid phase-field/cluster dynamics approach is capable of yielding highly accurate results, yet it is also computationally intensive. In the last decade, advances in data-driven methods such as deep learning provide an approach to develop surrogate FGR models that are much more computationally efficient than the hybrid model. Neural networks, once trained, can process new inputs without the use of specialized multiphysics tools. This paper introduces a physics-agnostic deep learning method based on convolutional neural networks (CNN) trained on the results from mesoscale simulations using the hybrid model to estimate instantaneous FGR flux from 2D microstructure images - a challenging problem, due to FGR dependency on complex spatial features in a microstructure such as the connectivity of fission gas bubbles to the free surface.

% \subsection{Convolutional Neural Networks}

CNNs are multilayer neural network architectures that infer features in grid-like input data through the application of convolutional operations, which can be construed as filters. The output of these filters represent the features, and the filter weights are trainable neural network parameters. Supervised learning uses training datasets comprised of input-output pairs which are fed into a gradient descent algorithm that finds the network parameters that minimize a loss function comparing the network output to the given training output when a particular input is applied to the network.

CNNs were originally developed in computer vision for image classification. However, CNNs can also be applied to solving regression problems for any gridlike data. For example, \cite{JongCheol2019} demonstrated the use of a CNN to infer cyanobacteria counts from hyperspectral images of up to 80 channels. The hybrid FGR model is solved on a finite element mesh, and therefore its output can be used to train a CNN. However, approximating a highly complex and spatially-dependent phenomenon such as FGR requires a comparably complex CNN architecture. For this task, one such CNN structure presents itself: DenseNet. 

% \subsection{DenseNet and Variants}

DenseNet \cite{Huang2017} is a class of CNN that preserves features from shallower network layers in deeper layers by appending the outputs of all previous layers onto the output of the current layer. Other CNN architectures such as ResNet \cite{He2015ResNet} have more localized interlayer connections, and therefore are less effective at preserving shallower features. Due to its compactness, improved performance, and reduced gradient vanishing with increased depth during training, DenseNet has become extensively utilized in many applications, including imaging \cite{Zhou2022}, superresolution \cite{Gunasekaran2023}, and remote sensing \cite{Guo2021}. To address the problem at hand, this study examines the following specific modifications to baseline DenseNet: intermediate regression layers for multiscale feature processing similar to \cite{Qin2020}, a combined spatial and channel attention mechanism \cite{Park2018}, and hybrid InceptionNet \cite{Zhang2019} blocks.

% --------------------
% \section{Objective}
% --------------------

The objective of this study is to train a DenseNet CNN to estimate the instantaneous FGR flux from 2D microstructure images. The training data is taken from hybrid FGR model simulations. The performance of four DenseNet variants equipped with intermediate regression layers are compared: baseline DenseNet, DenseNet with attention, and two DenseNets with attention and hybrid Inception blocks. The CNN is a less computationally expensive surrogate for the hybrid FGR model.

This paper is organized as follows. The methodology is discussed in Section \ref{sec:methodology}, including the CNN architecture, the training data, and the training and evaluation details. The results are presented in Section \ref{sec:results}. Section \ref{sec:conclusions} concludes the paper.
% --------------------
\section{Methodology} \label{sec:methodology}
% --------------------

\subsection{Neural Network Architecture}

\begin{figure*}[tbp]
    \centering
        \includegraphics[width=\textwidth]{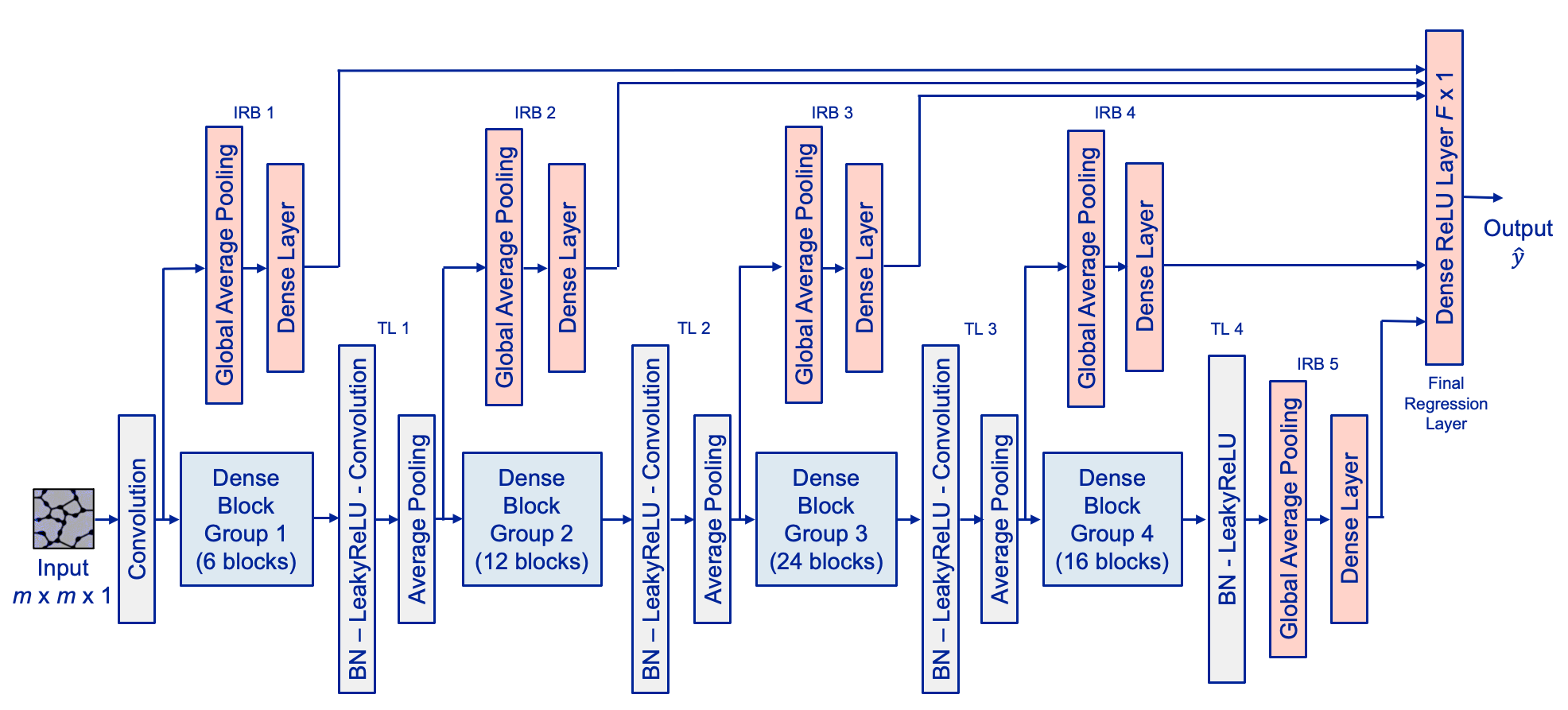}
    \caption{Schematic of the DenseNet architecture applied to predict instantaneous FGR flux from 2D microstructures.}
    \label{fig:1}
\end{figure*}

A high-level schematic of the proposed network architecture is shown in Fig.~\ref{fig:1}. Inspired by the DenseNet-121 architecture \cite{Huang2017}, the neural network consists of an initial convolutional layer followed by four dense block groups with associated transition layers (TL1 - TL4). Each dense block group is structured as a concatenation of a number of identical dense blocks, as indicated in Fig.~\ref{fig:1}. TL1 - TL3 consist of a batch normalization (BN) layer, followed by a leaky rectified linear unit (LeakyReLU) activation function layer and a convolutional layer. TL4 consists of only a BN layer, followed by a LeakyReLU layer. Compared to conventional CNNs, in which the output is obtained only from the final convolutional layer, the proposed architecture introduces a multiscale feature regression approach, where the output of the initial convolution layer and of each subsequent dense block group is passed to a fully connected (FC) intermediate regression block (IRB1 - IRB5). Each IRB consists of a global average pooling layer which produces a spatial average for each channel of the feature set, resulting in a 1D vector that is then inputted into a dense FC layer. The outputs of each intermediate regression block are concatenated and passed to a final regression layer with a rectified linear unit (ReLU) activation function that outputs the predicted instantaneous FGR flux $\hat{y}$ when the network is presented with an $m \times m \times 1$ image. In this manner, the CNN can better make predictions using both fine-scale and coarse-scale features. Intermediate classification layers have been successfully demonstrated on multiscale DenseNet variants for image classification problems \cite{Huang2018}, such as detecting lung cancer using fine-grained features \cite{Qin2020}, setting a precedent for their use in CNNs.

This study compares the performance of four dense block group structures - baseline DenseNet, DenseNet with attention, and two DenseNets with attention and hybrid Inception blocks - all using the same high-level framework shown in Fig.~\ref{fig:1}.

%\newpage

\begin{figure*}[tbp]
    \centering
        \includegraphics[width=\textwidth]{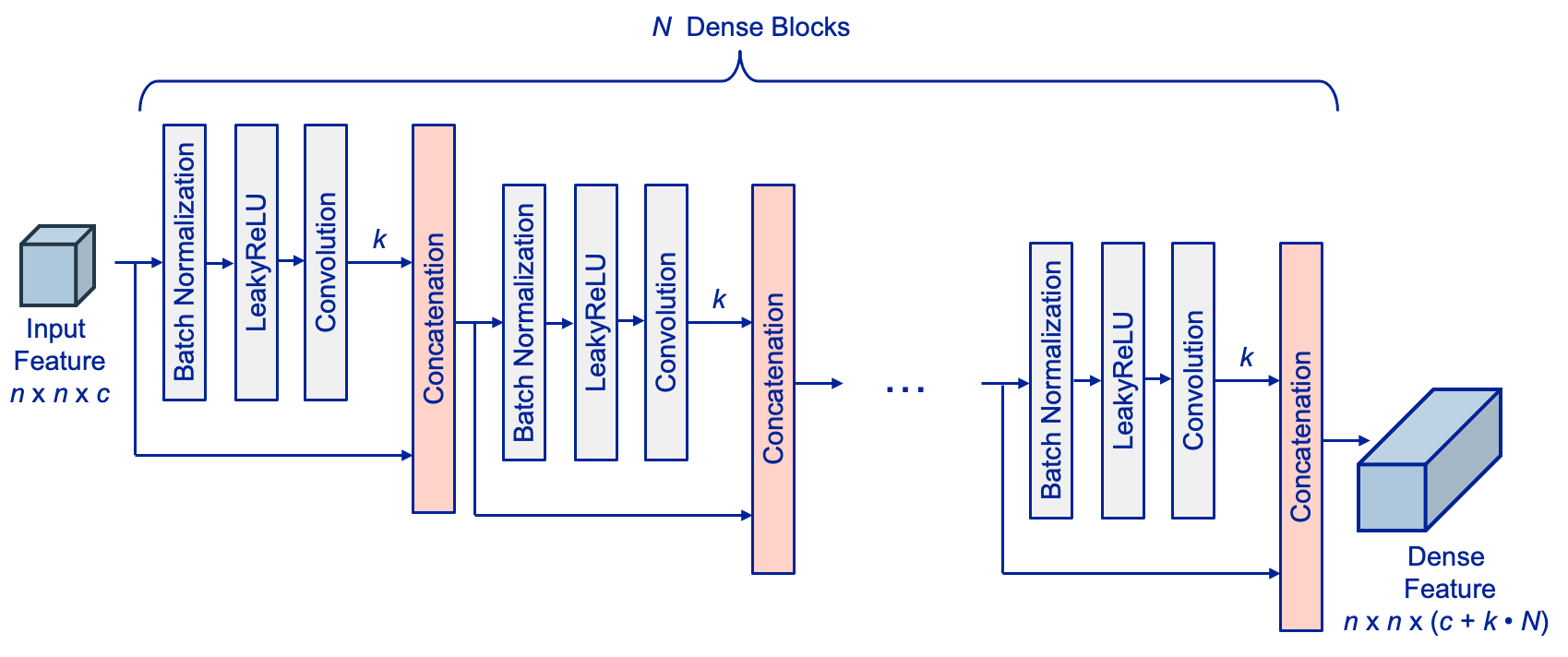}
    \caption{Schematic of the Baseline Dense Block Group - Basic DenseNet. There is no bottleneck nor compression.}
    \label{fig:2}
\end{figure*}

\paragraph{Baseline Dense Block Group - Basic DenseNet}

With reference to Fig.~\ref{fig:2}, a basic DenseNet block group consists of $N$ dense blocks, with each block consisting of a BN layer followed by a LeakyReLU layer and a convolutional layer. The output of the convolutional layer has $k$ channels, where $k$ is the growth rate of the network - the number of channels added to each layer in the network compared to the previous layer. The output of each dense block is concatenated with its input; this is the mechanism by which features from earlier in the network are passed to deeper layers. Comparing to the standard DenseNet-121 architecture, the proposed network does not use bottleneck and compression, in order to avoid loss of potentially useful features. Therefore, assuming that the size of the input feature is $n \times n \times c$, then the output of the basic dense block group with \textit{N} dense blocks has a size of $n \times n \times (c + k \cdot N)$.

\paragraph{Dense Attention Block Group}

\begin{figure*}[tbp]
    \centering
        \includegraphics[width=\textwidth]{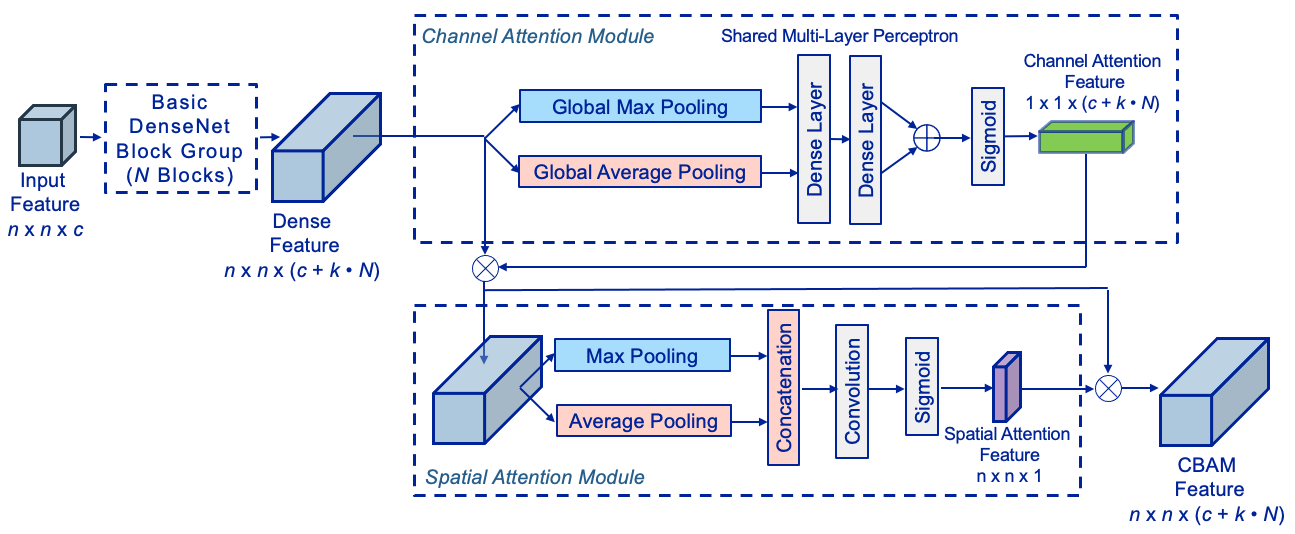}
    \caption{Schematic of the Dense Attention Block Group.}
    \label{fig:3}
\end{figure*}

Instantaneous FGR flux is highly dependent on spatial relationships in a microstructure such as fission gas bubble connectivity. While Baseline CNNs excel at detecting singular features, they struggle to capture these more complex relationships due to global pooling of feature maps that result in loss of spatial information. Attention mechanisms have been introduced to address this problem. Initially developed for time-series data \cite{Zhou2022}, attention mechanisms allow a network to dynamically focus on relevant features in each input image. The first notable CNN attention mechanism was the squeeze-and-excitation (SE) module \cite{Hu2017}, which only enabled channel attention. Illustrated in Fig. ~\ref{fig:3}, the Convolutional Block Attention Module (CBAM) \cite{JongCheol2019} combines an SE channel attention module with a spatial attention module. CBAM is attached to the output of a dense block. The CBAM-refined feature map has the same dimensionality as the output feature map of the dense block: $n \times n \times (c + k \cdot N)$.

\paragraph{Dense Inception Attention Block Groups}

\begin{figure*}[tbp]
    \centering
        \includegraphics[width=\textwidth]{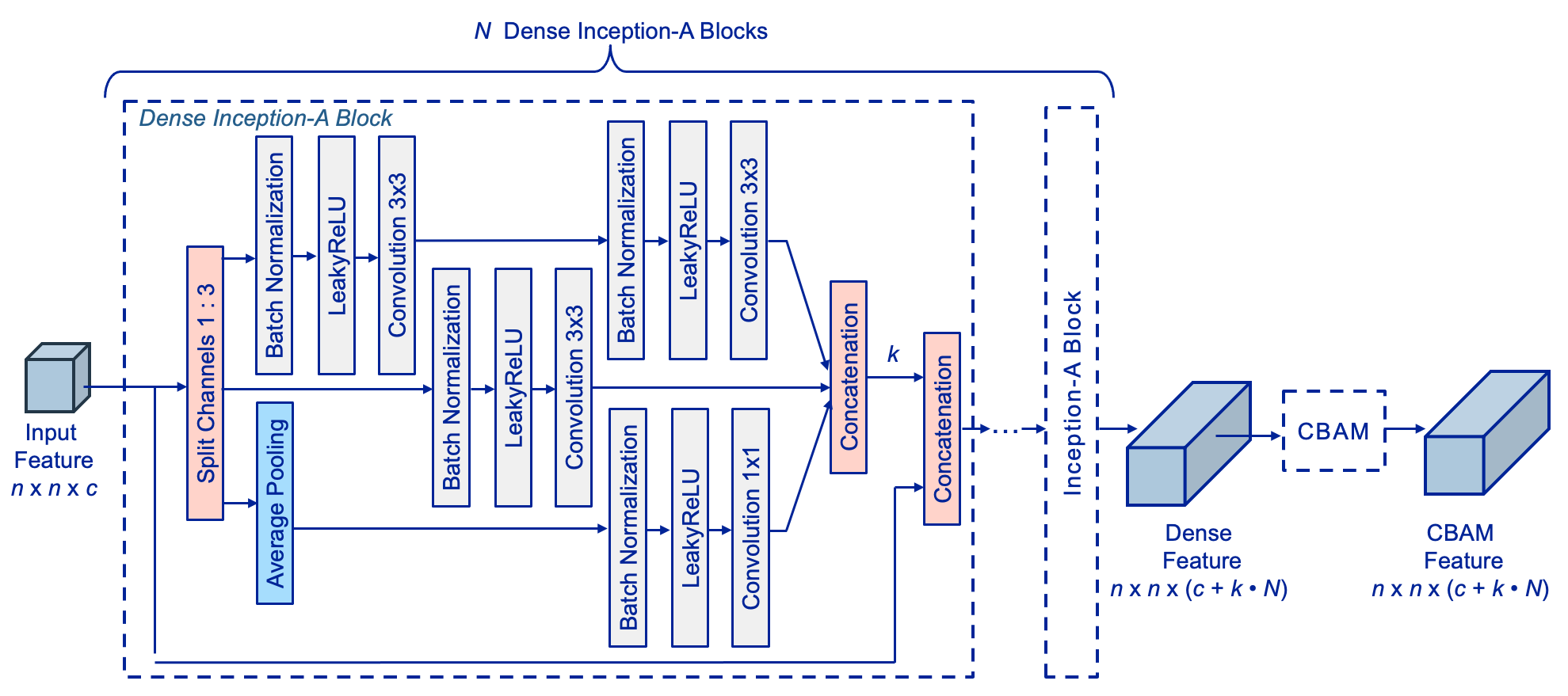}
    \caption{Schematic of the Dense Inception-A Attention Block Group.}
    \label{fig:4}
\end{figure*}

\begin{figure*}[tbp]
    \centering
        \includegraphics[width=\textwidth]{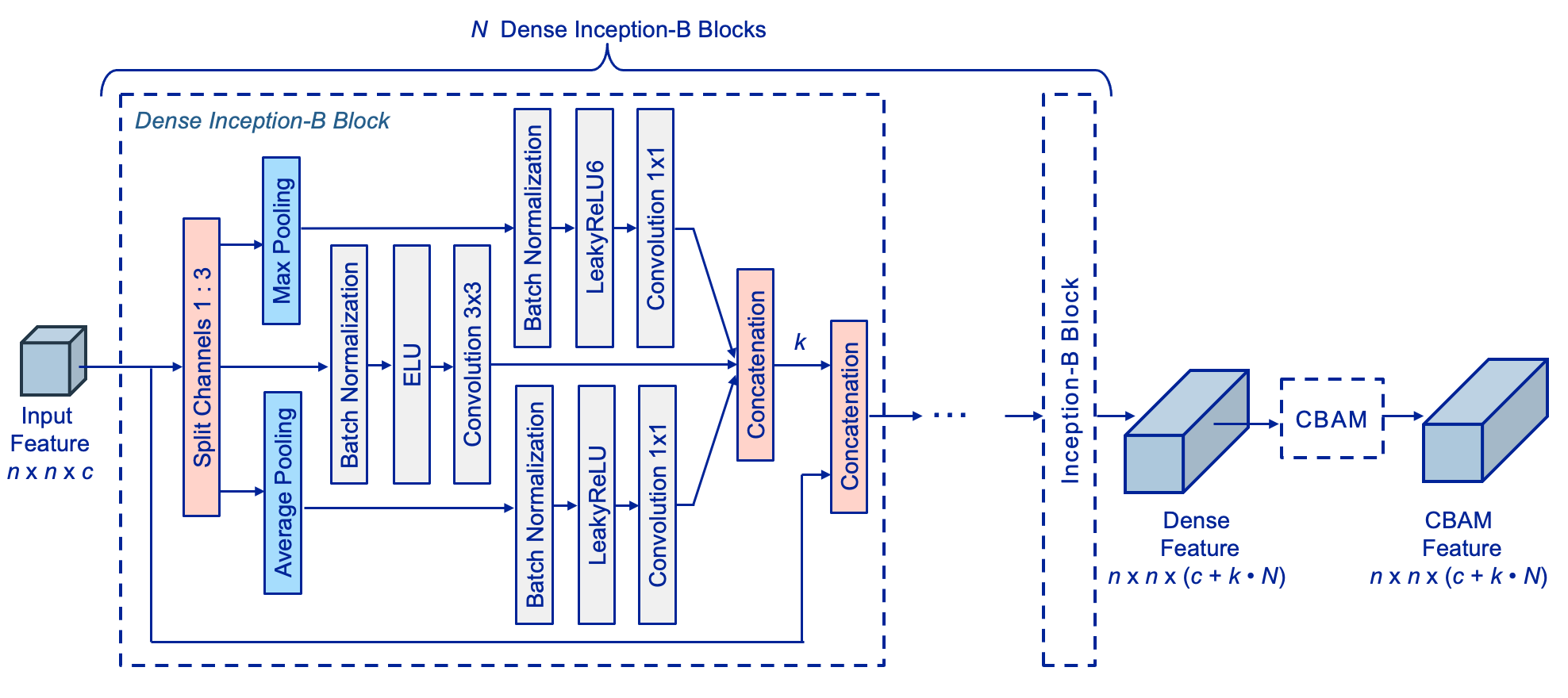}
    \caption{Schematic of the Dense Inception-B Attention Block Group.}
    \label{fig:5}
\end{figure*}

The multiscale detection abilities of a DenseNet can be further enhanced by replacing the standard single-convolution layer in a DenseNet block with a set of layers based on the InceptionNet architecture, as described in \cite{Zhang2019} - see Fig. ~\ref{fig:4} and Fig. ~\ref{fig:5}. Dense Inception blocks are able to both reduce overfitting and improve a network's fine-grained feature detection ability - useful for capturing a microstructure's complex geometry.

A Dense Inception block divides the input channel-wise into three branches and performs different combinations of pooling, convolution, and activation function operations on each branch, allowing the block to detect a significantly larger range of features across different scales while reducing the overall number of required trainable parameters. Inspired by \cite{Zhang2019}, we consider two Inception block architectures. Inception-A has a branch with a cascade of two $3 \times 3$ convolution layers and an average pooling branch, while Inception-B replaces the cascade with a $3 \times 3 \times 3$ convolution branch with a max pooling layer followed by a $1\times1\times1$ convolution layer. All branches in Inception-A use the same LeakyReLU activation function, while Inception-B uses three different activation functions: LeakyReLU, a leaky rectified linear unit activation with upper limit of $6.0$ (LeakyReLU6), and an exponential linear unit activation function (ELU).

\newpage

Finally, a CBAM block can be attached to the output of a Dense Inception block group to gain the benefits of attention mechanisms. In this study, we do not use a bottleneck or compression inside a Dense Inception block. Therefore, the output feature map of a Dense Inception Attention block group retains the same dimensionality as that of a Dense Attention block group, assuming both have \textit{N} blocks.

\subsection{Network Architecture Parameters}

The network architecture parameters chosen for the present study are summarized, below. 

\begin{enumerate}
  \item \textbf{Image input} -- $4\times4$ zero padding is applied to the input images for the network to better detect the surfaces of the microstructure.
  \item \textbf{Initial convolution layer} (see Fig.~\ref{fig:1}) -- $7\times7$ kernel, stride = 2, padding same as image input, and \textit{2k} channels (where \textit{k} is the growth rate). Using the same padding maintains the input's spatial dimensions in the output.
  \item \textbf{Transition layers} (see Fig.~\ref{fig:1}) -- 1x1 convolution, stride = 1, the same padding, followed by $2\times2$ average pooling with stride = 2 and the padding.
  \item \textbf{Convolution layers in baseline DenseNet block groups} (see Fig. 2) -- $3\times3$ kernel, stride = 1, the same padding, and \textit{k} channels.
  \item \textbf{Convolution layer in spatial attention module from CBAM} (see Fig. 3) -- $7\times7$ kernel, stride = 1, same padding, and 1 channel. 
  \item \textbf{Dense Inception-A block configuration} (see Fig. 4) -- The first branch uses two $3\times3$ convolution layers, stride = 1, and same padding. The second branch uses a $3\times3$ convolution layer, stride = 1 and the same padding. The third branch uses a $3\times3$ average pooling layer, stride = 1, same padding, and a $1\times1$ convolution layer with stride = 1 and the same padding.
  \item \textbf{Dense Inception-B block configuration} (see Fig. 5) - The first branch uses stride = 1, $3\times3$ max pooling layer, and a $1\times1$ convolution layer with stride = 1 and the same padding. The second branch uses a 3x3 convolution layer, stride = 1 and the same padding. The third branch used a $3\times3$ average pooling layer, stride = 1, the same padding, and a $1\times1$ convolution layer with stride = 1 and same padding.
  \item \textbf{Activation functions} -- Except for the final layer and where otherwise noted, LeakyReLU activation (\(\alpha = 0.2\)) is used throughout the network to prevent neuron ``death.'' If a high gradient is applied to a neuron using conventional ReLU activation, the neuron could be set to a very small value and not be able to recover, i.e.\ ``die.'' LeakyReLU enables the reactivation of ``dead'' neurons. The max pooling branch inside the Dense Inception-B block (see Fig. 5) uses LeakyReLU6 (\(\alpha = 0.2\)), which is simply LeakyReLU with a maximum value of 6. The Inception-B $3\times3$ convolution branch uses ELU activation with \(\alpha = 0.1\).
  \item \textbf{Final regression layer} -- The final FC layer uses ReLU activation with a single output neuron to output the final instantaneous FGR flux prediction, ensuring that the network does not output physically impossible negative values.
  \item \textbf{Attention ratio} -- \(r = 8\) used for networks with CBAM (see Fig. 3) in the shared MLP as part of the squeeze mechanism specific to the channel attention module.
  \item \textbf{Growth rate} -- \(k = 32\) used for  baseline DenseNet and Dense Attention architectures. \(k = 33\) used for the Dense Inception Attention architecture, as \textit{k} must be divisible by 3 in this case.
\end{enumerate}

Table ~\ref{tab:1} summarizes the feature map dimensions throughout the networks examined in this study, assuming input image dimensions of $101 \times 101 \times 1$ ($109 \times 109 \times 1$ after $4\times4$ zero padding). Notably, both Dense Inception + CBAM configurations have approximately half the trainable parameters of both baseline DenseNet and DenseNet + CBAM, pointing to a more efficient architecture. The size of the final regression layer's input vector is 4,288 for the baseline DenseNet and DenseNet + CBAM networks, and 4,422 for the Dense Inception + CBAM networks.

\begin{table}[tbph]
\caption{Summary of feature map dimensions throughout examined networks.}
\label{tab:1}
\centering
\begin{tabular}{cccc|}
\cline{2-4}
\multicolumn{1}{l|}{} &
  \multicolumn{1}{c|}{\textbf{\begin{tabular}[c]{@{}c@{}}DenseNet\\ (k=32)\end{tabular}}} &
  \multicolumn{1}{c|}{\textbf{\begin{tabular}[c]{@{}c@{}}DenseNet\\+ CBAM (k=32)\end{tabular}}} &
  \textbf{\begin{tabular}[c]{@{}c@{}}Inception A/B\\ + CBAM (k=33)\end{tabular}} \\ \hline
\multicolumn{1}{|c|}{\textbf{Blocks/Layers}} &
  \multicolumn{1}{c|}{\textbf{Output Size}} &
  \multicolumn{1}{c|}{\textbf{Output Size}} &
  \textbf{Output Size} \\ \hline
\multicolumn{1}{|c|}{Convolution}                     & \multicolumn{1}{c|}{$55 \times 55 \times 64$}   & \multicolumn{1}{c|}{$55 \times 55 \times 64$}   & $55 \times 55 \times 66$   \\ \hline
\multicolumn{1}{|c|}{Intermediate Regression Block 1} & \multicolumn{1}{c|}{64}             & \multicolumn{1}{c|}{64}             & 66             \\ \hline
\multicolumn{4}{|l|}{\textit{Dense Block Group 1}}                                                                                                 \\ \hline
\multicolumn{1}{|c|}{Basic DenseNet Block $\times 6$}        & \multicolumn{1}{c|}{$55 \times 55 \times 256$}  & \multicolumn{1}{c|}{$55 \times 55 \times 256$}  & N/A            \\ \hline
\multicolumn{1}{|c|}{Dense Inception Block $\times 6$}       & \multicolumn{1}{c|}{N/A}            & \multicolumn{1}{c|}{N/A}            & $55 \times 55 \times 264$  \\ \hline
\multicolumn{1}{|c|}{CBAM}                            & \multicolumn{1}{c|}{N/A}            & \multicolumn{1}{c|}{$55 \times 55 \times 256$}  & $55 \times 55 \times 264$  \\ \hline
\multicolumn{1}{|c|}{Transition Layer 1}              & \multicolumn{1}{c|}{$27 \times 27 \times 256$}  & \multicolumn{1}{c|}{$27 \times 27 \times 256$}  & $27 \times 27 \times 264$  \\ \hline
\multicolumn{1}{|c|}{Intermediate Regression Block 2} & \multicolumn{1}{c|}{256}            & \multicolumn{1}{c|}{256}            & 264            \\ \hline
\multicolumn{4}{|l|}{\textit{Dense Block Group 2}}                                                                                                 \\ \hline
\multicolumn{1}{|c|}{Basic DenseNet Block $\times 12$}       & \multicolumn{1}{c|}{$27 \times 27 \times 640$}  & \multicolumn{1}{c|}{$27 \times 27 \times 640$}  & N/A            \\ \hline
\multicolumn{1}{|c|}{Dense Inception Block $\times 12$}      & \multicolumn{1}{c|}{N/A}            & \multicolumn{1}{c|}{N/A}            & $27 \times 27 \times 660$  \\ \hline
\multicolumn{1}{|c|}{CBAM}                            & \multicolumn{1}{c|}{N/A}            & \multicolumn{1}{c|}{$27 \times 27 \times 640$}  & $27 \times 27 \times 660$  \\ \hline
\multicolumn{1}{|c|}{Transition Layer 2}              & \multicolumn{1}{c|}{$13 \times 13 \times 640$}  & \multicolumn{1}{c|}{$13 \times 13 \times 640$}  & $13 \times 13 \times 660$  \\ \hline
\multicolumn{1}{|c|}{Intermediate Regression Block 3} & \multicolumn{1}{c|}{640}            & \multicolumn{1}{c|}{640}            & 660            \\ \hline
\multicolumn{4}{|l|}{\textit{Dense Block Group 3}}                                                                                                 \\ \hline
\multicolumn{1}{|c|}{Basic DenseNet Block $\times 24$}       & \multicolumn{1}{c|}{$13 \times 13 \times 1408$} & \multicolumn{1}{c|}{$13 \times 13 \times 1408$} & N/A            \\ \hline
\multicolumn{1}{|c|}{Dense Inception Block $\times 24$}      & \multicolumn{1}{c|}{N/A}            & \multicolumn{1}{c|}{N/A}            & $13 \times 13 \times 1452$ \\ \hline
\multicolumn{1}{|c|}{CBAM}                            & \multicolumn{1}{c|}{N/A}            & \multicolumn{1}{c|}{$13 \times 13 \times 1408$} & $13 \times 13 \times 1452$ \\ \hline
\multicolumn{1}{|c|}{Transition Layer 3}              & \multicolumn{1}{c|}{$6 \times 6 \times 1408$}   & \multicolumn{1}{c|}{$6 \times 6 \times 1408$}   & $6 \times 6 \times 1452$   \\ \hline
\multicolumn{1}{|c|}{Intermediate Regression Block 4} & \multicolumn{1}{c|}{1408}           & \multicolumn{1}{c|}{1408}           & 1452           \\ \hline
\multicolumn{4}{|l|}{\textit{Dense Block Group 4}}                                                                                                 \\ \hline
\multicolumn{1}{|c|}{Basic DenseNet Block $\times 16$}       & \multicolumn{1}{c|}{$6 \times 6 \times 1920$}   & \multicolumn{1}{c|}{$6 \times 6 \times 1920$}   & N/A            \\ \hline
\multicolumn{1}{|c|}{Dense Inception Block $\times 16$}      & \multicolumn{1}{c|}{N/A}            & \multicolumn{1}{c|}{N/A}            & $6 \times 6 \times 1980$   \\ \hline
\multicolumn{1}{|c|}{CBAM}                            & \multicolumn{1}{c|}{N/A}            & \multicolumn{1}{c|}{$6 \times 6 \times 1920$}   & $6 \times 6 \times 1980$   \\ \hline
\multicolumn{1}{|c|}{Transition Layer 4}              & \multicolumn{1}{c|}{$6 \times 6 \times 1920$}   & \multicolumn{1}{c|}{$6 \times 6 \times 1920$}   & $6 \times 6 \times 1980$   \\ \hline
\multicolumn{1}{|c|}{Intermediate Regression Block 5} & \multicolumn{1}{c|}{1920}           & \multicolumn{1}{c|}{1920}           & 1980           \\ \hline
\multicolumn{1}{|c|}{Dense Linear Layer}              & \multicolumn{1}{c|}{1}              & \multicolumn{1}{c|}{1}              & 1              \\ \hline
\multicolumn{1}{|c|}{Number of Trainable Parameters}  & \multicolumn{1}{c|}{25,042,177}     & \multicolumn{1}{c|}{26,583,331}     & \begin{tabular}{@{}c@{}} 15,055,569 (A) \\ 13,278,739 (B) \end{tabular}    \\ \hline
\end{tabular}
\end{table}

\subsection{Training and Evaluation Data}

As FGR and gas bubble growth inside fuel rods occur during the active operation of a nuclear reactor, it is very difficult to obtain in situ data characterizing the evolving bubble structures. It is more practical to generate synthetic data that includes both the microstructure and FGR in the high volumes required for neural network training. This study uses the hybrid phase field/cluster dynamics model of fission gas behavior developed by Kim et al.~\cite{Tonks2022}. The phase field model is implemented in the mesoscale MARMOT code based on the open-source Multiphysics Object-Oriented Simulation Environment (MOOSE) \cite{lindsay2022moose}. The cluster dynamics model is implemented in the Xolotl library \cite{xolotl2022}. The two codes are coupled using the MultiApp capability available in MOOSE \cite{gaston2015physics}. For more detail about the fission gas model, see the paper by Kim et al.~\cite{Tonks2022}. The hybrid model has been modified to include fast grain boundary and surface diffusion and a free surface to include FGR \cite{muntaha2023}. 

To generate the training data, we apply the modified hybrid fission gas model \cite{muntaha2023} to simulate the microstructure evolution and the flux of fission gas from the left surface in 2D 10-grain microstructures for 365 days at 1500 K. The simulations are carried out in a 15 $\mu$m by 15 $\mu$m domain with 20 initial 480 nm radius fission gas bubbles. 100 elements are used in both the $x$- and $y$-directions, resulting in a mesh size of 150 nm. The simulation uses an adaptive timestepping scheme with an initial timestep of 10 seconds; each simulation uses up to 700 time steps and the instantaneous flux of fission gas from the left surface is calculated for the microstructure at each time step. Zero flux boundary conditions are applied on the right boundary and periodic boundary conditions on the top and bottom boundaries. The simulation is repeated for 100 different initial grain boundary and fission gas bubble structures, resulting in over 72,000 microstructure images and corresponding instantaneous FGR flux values. 2D simulations are used here to reduce the computational cost of generating the training data.

\begin{figure}
    \centering
        \includegraphics[width=3in]{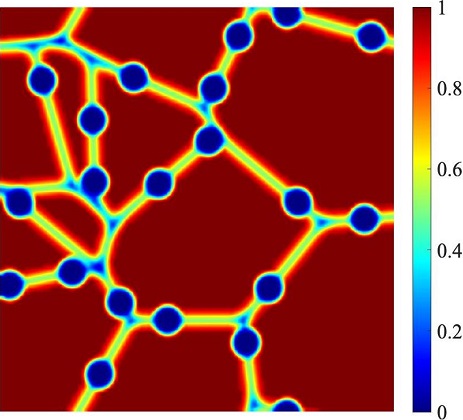}
    \caption{Microstructure image obtained from the hybrid FGR model. Red indicates UO$_2$, blue a void, and all other colors represent grain boundaries or void surfaces.}
    \label{fig:6}
\end{figure}

Each microstructural image consists of a set of floating point values ranging between 0 and 1, corresponding to the phase parameter of the microstructure at each grid point. A value of 0 represents a void, a value of 1 represents UO$_2$ nuclear fuel, and a value between 0 and 1 represents a grain boundary or void surface. In order to improve computational efficiency, MARMOT uses automatic mesh adaptivity to automatically coarsen or refine different regions of a simulation mesh, simulating low-error regions such as the interiors of grains with less fidelity than high-error regions such as grain boundaries. Because of this, the raw phase parameter image data output by MARMOT does not correspond to a regular 2D grid, and therefore it is necessary to interpolate each image onto a $101 \times 101 \times 1$ grid during preprocessing. Fig. ~\ref{fig:6} shows an example of an input microstructure image.

To minimize floating point errors, the FGR flux values of the original data are scaled up by a factor of 1000 for training, then scaled back to the original range for evaluation. The instantaneous FGR value of a microstructure does not change if it is mirrored across the axis perpendicular to the free surface on the left, allowing for doubling of the dataset to approximately 144,000 images by adding, for each image, its vertically-mirrored version with the same instantaneous FGR flux. 

\subsection{Training and Evaluation Details}

The specified network architectures are implemented, trained, and evaluated in Python 3, using the Tensorflow 2.7.0 framework with Keras \cite{tensorflow2015-whitepaper}. The code is executed on the University of Florida's HiPerGator supercomputer \cite{ufrc} on a partition with two Intel Xeon Gold 6142 CPUs @ 2.60 GHz and one NVIDIA A100 GPU. On this configuration, training takes approximately 18-24 hours, depending on the neural network architecture.

For training, mean absolute error (MAE), standard for regression problems, is used as the loss function for both training and testing. As the absolute instantaneous FGR flux output values are scaled to arbitrary units, mean absolute percentage error (MAPE) is used as the primary evaluation metric. Stochastic Gradient Descent is used as the optimizer, as in the original DenseNet paper \cite{Huang2017}, with a fixed learning rate of \(\alpha = 0.001\). L2 weight regularization with weight decay = $10^{-4}$ is used in all convolutional layers. All weights are initialized using the He method \cite{He2015}. Each network is trained for a total of 100 epochs, with a batch size of 32 images.

This study uses a 90\%/5\%/5\% training/evaluation/test split of the data set, with performance on the test set reported as final results. The input data is shuffled randomly to prevent spurious sequential effects. Linear regression analysis is performed using the MATLAB \cite{MATLAB} \textit{fitlm} function to calculate the R-squared value of predicted vs.\ simulated instantaneous FGR flux values for each tested network, with values closer to 1 indicating greater network predictive power. For interpretability, saliency maps are generated using a Tensorflow \cite{tensorflow2015-whitepaper} GradientTape-enabled back-propagation visualization method \cite{Simonyan2014}. These maps provide a visual representation of the regions of an input image that activate the network, indicating which regions of the microstructure have the largest impact on the FGR.

% --------------------
\section{Results} \label{sec:results}
% --------------------

\begin{figure*}[tbp]
    \centering
        \includegraphics[scale=0.47]{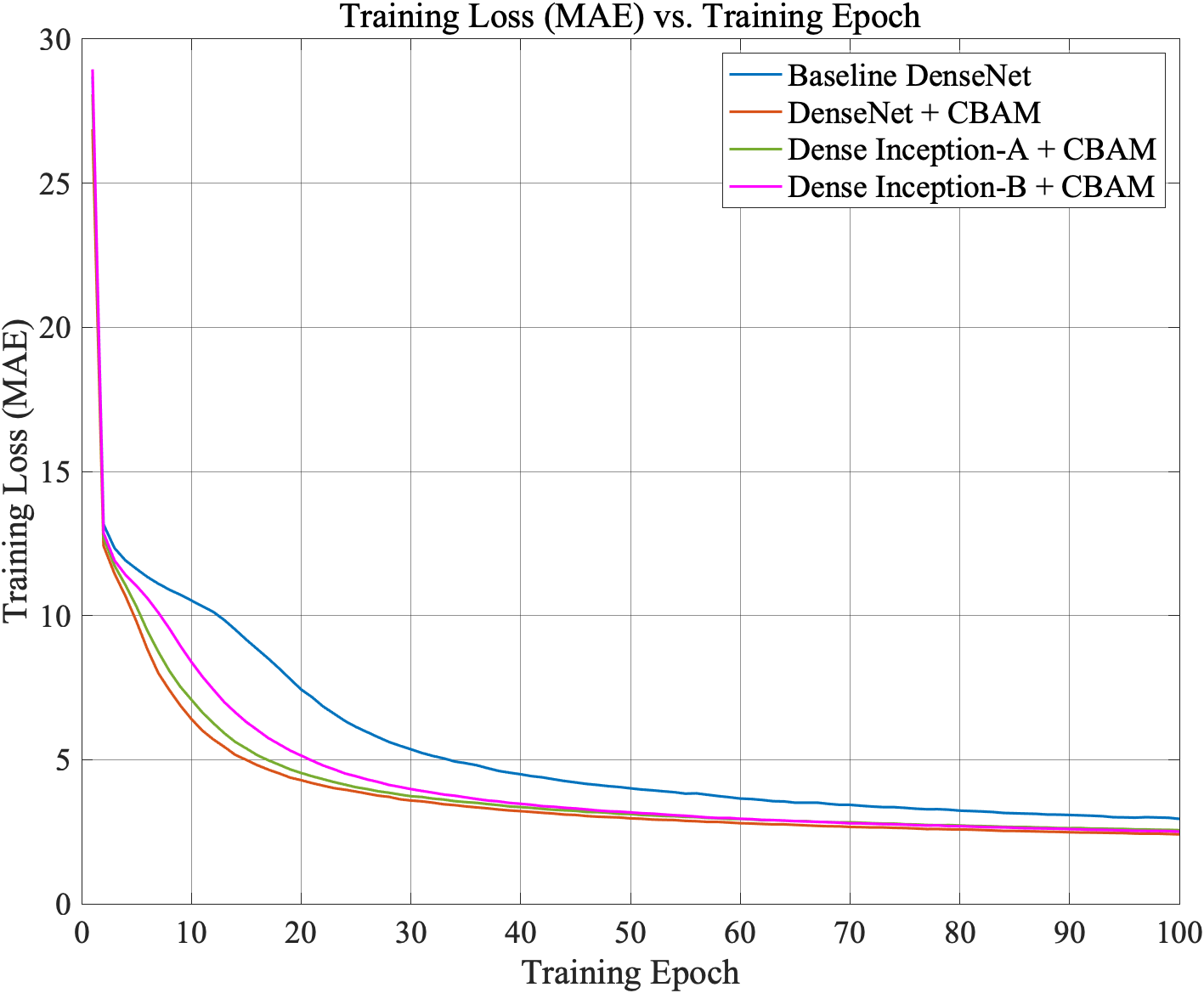}
    \caption{Plot of MAE training loss vs.\ training epoch for each examined network.}
    \label{fig:7}
\end{figure*}

Figure~\ref{fig:7} shows a plot of the MAE training loss vs.\ training epoch for each examined network. Baseline DenseNet underperforms compared to other architectures for all epochs. After 40 epochs, all the attention-enabled networks have reached comparable loss values. The DenseNet Attention (DenseNet + CBAM) network without Inception blocks generates the lowest MAE training loss for all epochs, but this could also be a sign of overfitting, as will be discussed later in this section.

\begin{figure*}[t]
\centering
\begin{subfigure}{0.48\textwidth}
    \centering
    \includegraphics[width=\linewidth]{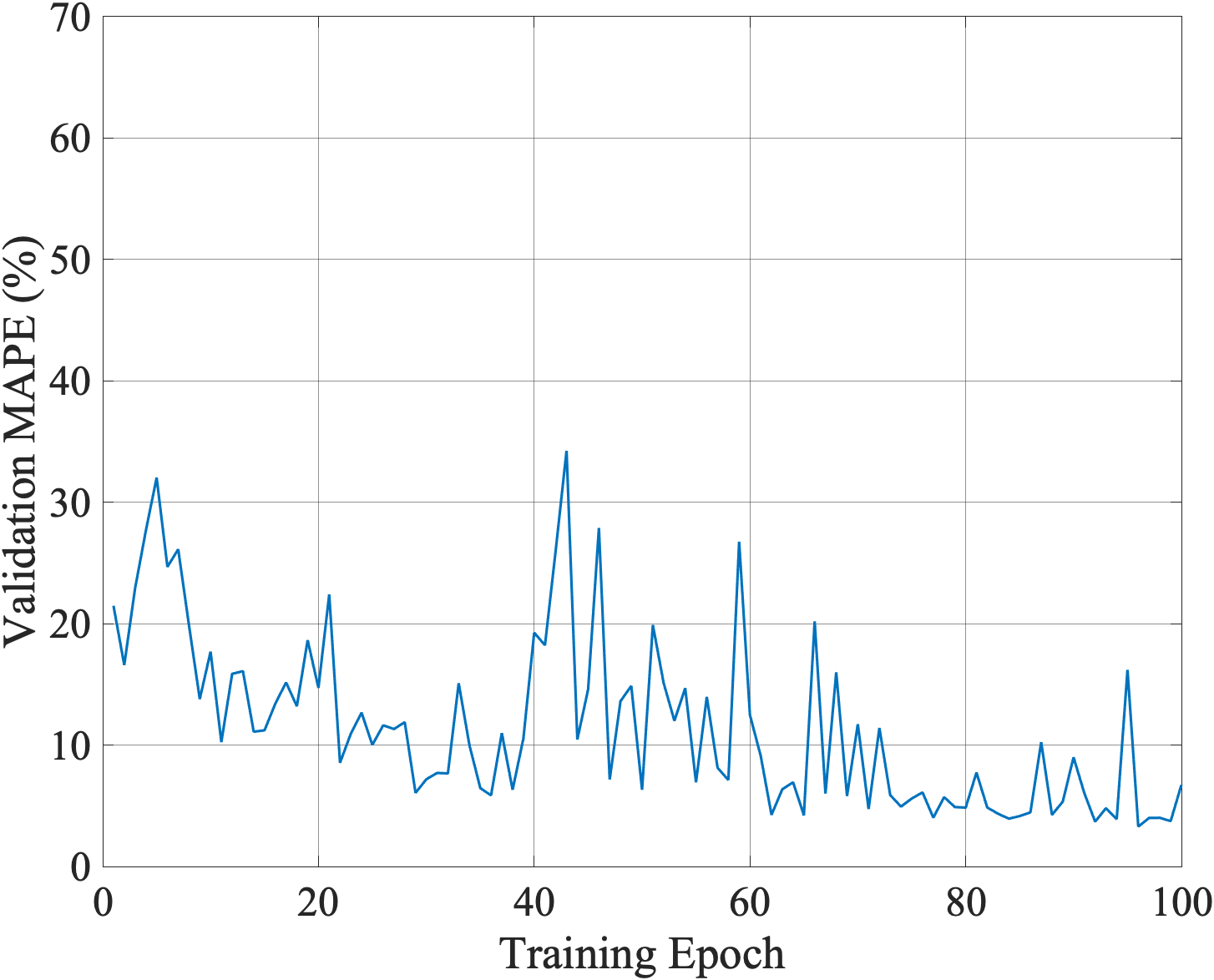}
    \caption{Baseline DenseNet}
\end{subfigure}
\begin{subfigure}{0.48\textwidth}
    \centering
    \includegraphics[width=\linewidth]{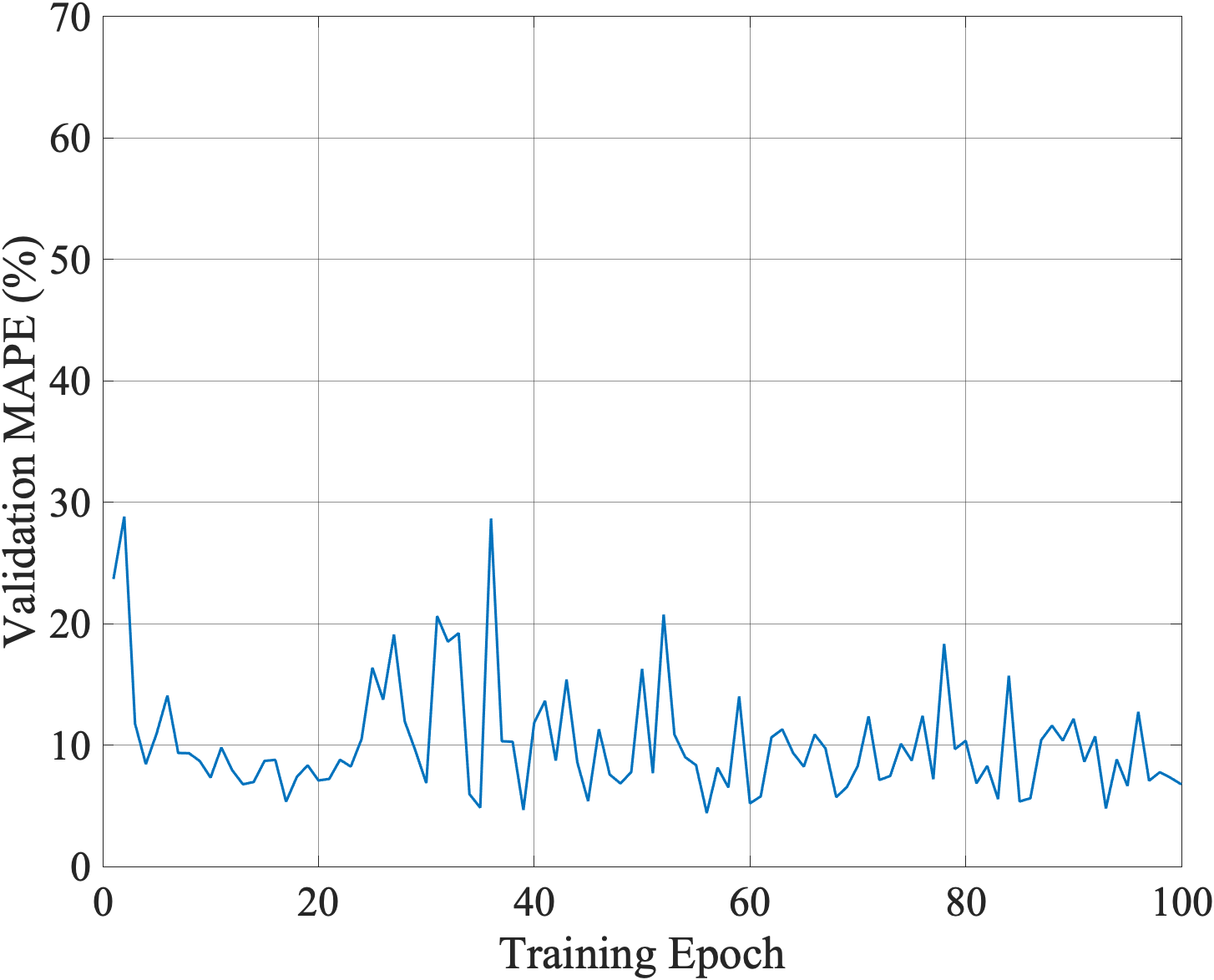}
    \caption{DenseNet + CBAM}
\end{subfigure}
\begin{subfigure}{0.48\textwidth}
    \centering
    \includegraphics[width=\linewidth]{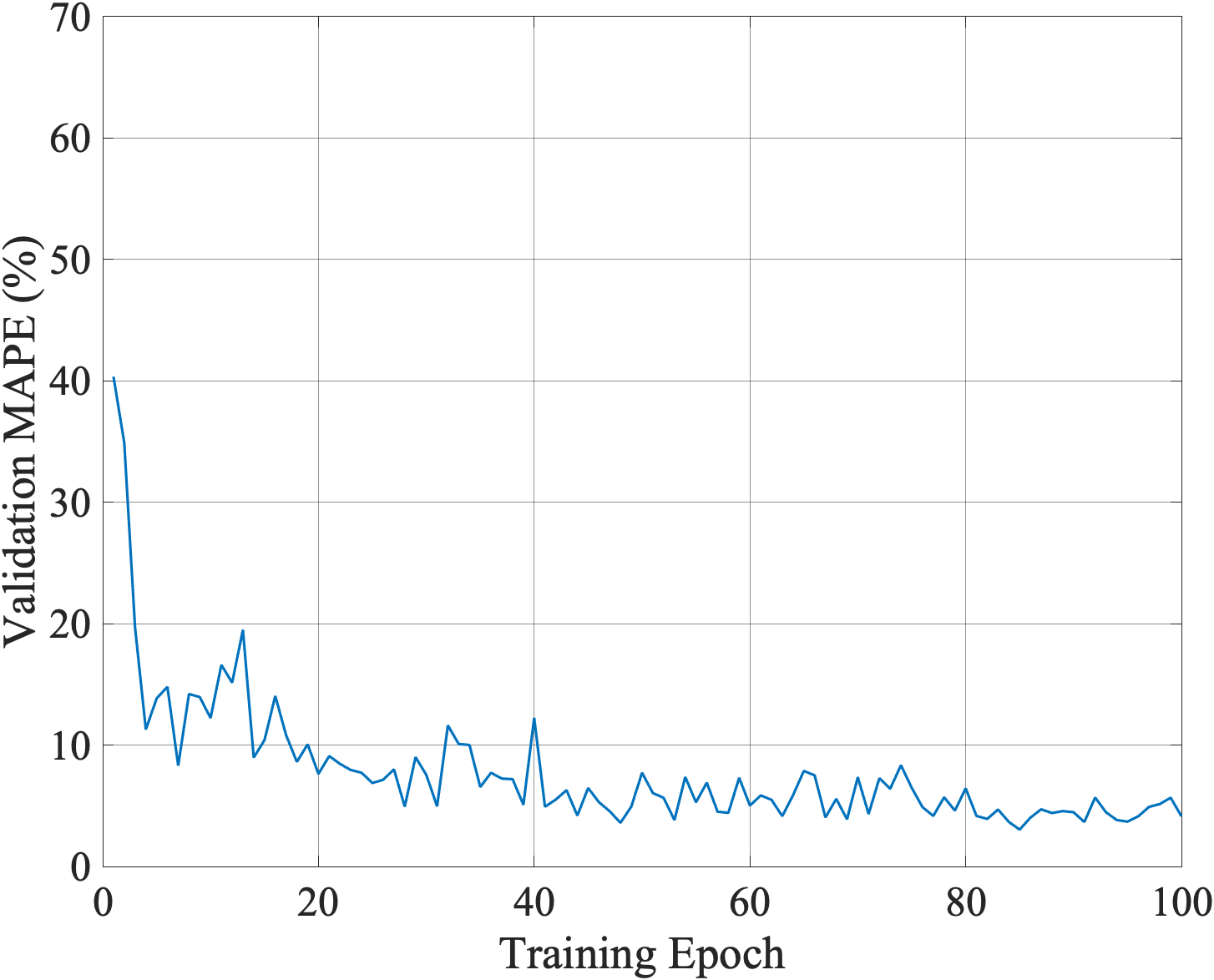}
    \caption{DenseNet Inception-A + CBAM}
\end{subfigure}
\begin{subfigure}{0.48\textwidth}
    \centering
    \includegraphics[width=\linewidth]{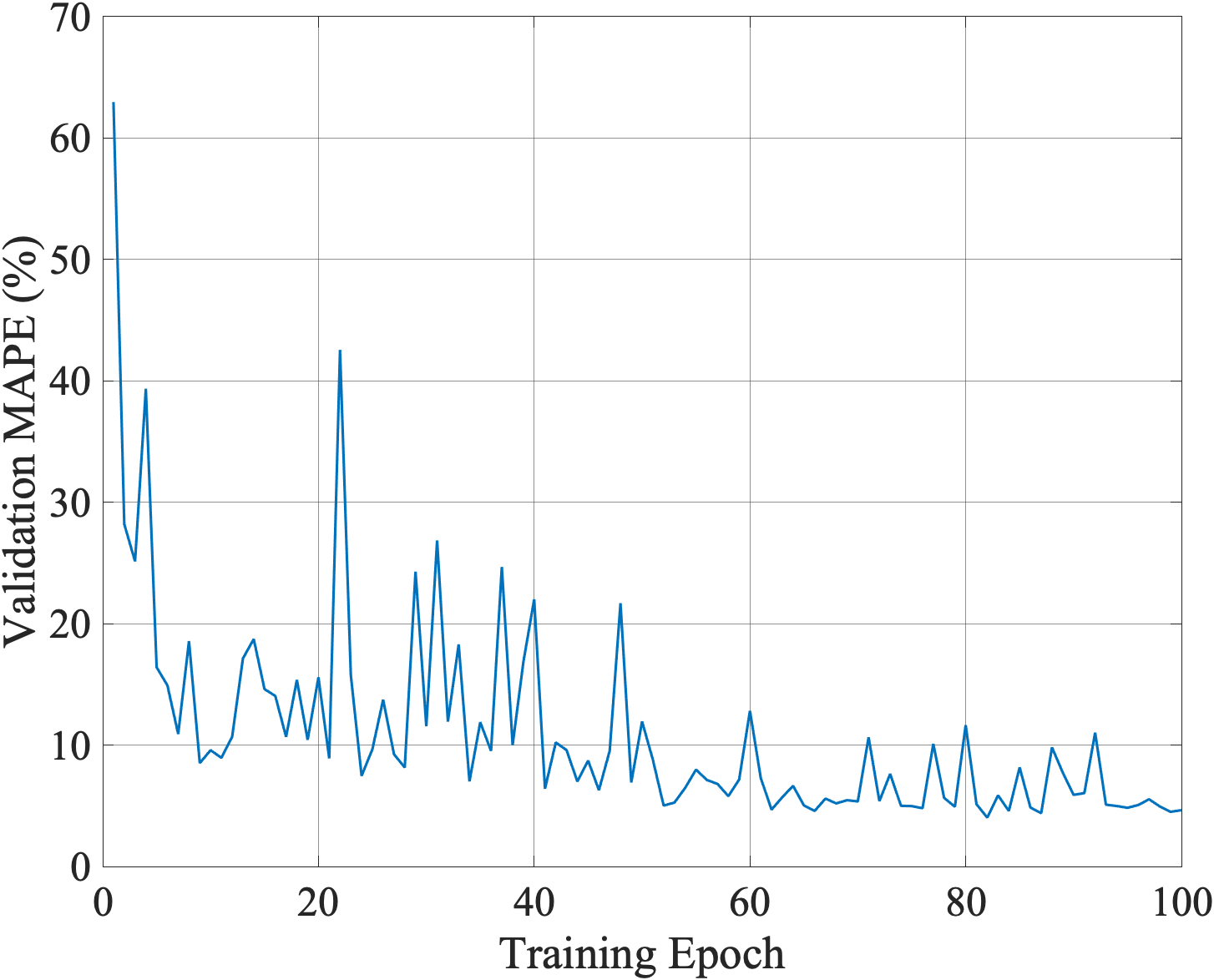}
    \caption{DenseNet Inception-B + CBAM}
\end{subfigure}
\caption{Plot of validation MAPE vs.\ training epoch for each examined network.}
\label{fig:8}
\end{figure*}

Figure~\ref{fig:8} compares the validation MAPE vs.\ training epoch behavior across the four tested networks. DenseNet Inception-A + CBAM appears to exhibit better training stability from epoch to epoch, judging from reduced fluctuations in validation MAPE during the training process.

%\newpage

Table ~\ref{tab:2} summarizes the training and evaluation performance of the examined DenseNet architectures. DenseNet + CBAM without Inception shows the lowest training MAPE (2.88\%), yet also shows the highest validation MAPE (6.75\%) - an indicator of overfitting to the training set. The Dense Inception + CBAM networks shows comparable training and validation MAPE values, with Dense Inception-A + CBAM providing the lowest validation MAPE (4.13\%). This behavior points to greater generalizability on the part of the Inception blocks, which is the intended outcome of these architectures. Dense Inception-A + CBAM also provides the lowest MAPE on the test set (4.40\%).

To examine the performance of the examined networks for very low instantaneous FGR flux values, we calculate an adjusted test MAPE by filtering out all input images with associated instantaneous FGR flux values of less than 0.001 (unscaled), with only 47 out of 7,227 test data points fitting this criterion. Based on the difference between the test MAPE and the adjusted test MAPE, Dense Inception-A + CBAM appears to be the most robust configuration with respect to very low instantaneous FGR flux values, followed by Dense Inception-B + CBAM. DenseNet + CBAM without Inception appears to be the least robust architecture with respect to this criterion.

%\newpage

\begin{table}[!tbph]
\centering
  \caption{Comparative training and evaluation performance of examined DenseNet architectures.}
  \label{tab:2}
\begin{tabular}{lllll|}
\cline{2-5}
\multicolumn{1}{l|}{} &
  \multicolumn{1}{c|}{\textbf{\begin{tabular}[c]{@{}c@{}}Baseline \\ DenseNet\end{tabular}}} &
  \multicolumn{1}{c|}{\textbf{\begin{tabular}[c]{@{}c@{}}DenseNet \\ + CBAM\end{tabular}}} &
  \multicolumn{1}{c|}{\textbf{\begin{tabular}[c]{@{}c@{}}Inception-A \\ + CBAM\end{tabular}}} &
  \multicolumn{1}{c|}{\textbf{\begin{tabular}[c]{@{}c@{}}Inception-B\\ + CBAM\end{tabular}}} \\ \hline
\multicolumn{5}{|l|}{\textit{Training and Validation Data Set Performance @ 100 Training   Epochs}}                                                           \\ \hline
\multicolumn{1}{|l|}{\textbf{Training Loss (MAE)}}     & \multicolumn{1}{r|}{2.94}    & \multicolumn{1}{r|}{2.40}    & \multicolumn{1}{r|}{2.55}    & \multicolumn{1}{r|}{2.50}    \\ \hline
\multicolumn{1}{|l|}{\textbf{Training MAPE (\%)}}      & \multicolumn{1}{r|}{4.31}  & \multicolumn{1}{r|}{2.88}  & \multicolumn{1}{r|}{3.05}  & \multicolumn{1}{r|}{3.23}  \\ \hline
\multicolumn{1}{|l|}{\textbf{Validation Loss (MSE)}}   & \multicolumn{1}{r|}{4.00}    & \multicolumn{1}{r|}{3.59}    & \multicolumn{1}{r|}{3.83}    & \multicolumn{1}{r|}{3.61}    \\ \hline
\multicolumn{1}{|l|}{\textbf{Validation MAPE (\%)}}    & \multicolumn{1}{r|}{6.72}  & \multicolumn{1}{r|}{6.75}  & \multicolumn{1}{r|}{4.13}  & \multicolumn{1}{r|}{4.63}  \\ \hline
\multicolumn{5}{|l|}{\textit{Test Data Set Performance}}                                                                                                      \\ \hline
\multicolumn{1}{|l|}{\textbf{Test MAPE (\%)}}          & \multicolumn{1}{r|}{8.56}  & \multicolumn{1}{r|}{10.85} & \multicolumn{1}{r|}{4.40}  & \multicolumn{1}{r|}{5.14}  \\ \hline
\multicolumn{1}{|l|}{\textbf{Adjusted Test MAPE (\%)}} & \multicolumn{1}{r|}{5.17}  & \multicolumn{1}{r|}{3.52}  & \multicolumn{1}{r|}{3.56}  & \multicolumn{1}{r|}{4.00}  \\ \hline
\multicolumn{1}{|l|}{\textbf{$R^{2}$ (\%)}}          & \multicolumn{1}{r|}{98.87} & \multicolumn{1}{r|}{99.23} & \multicolumn{1}{r|}{99.10} & \multicolumn{1}{r|}{98.88} \\ \hline
\multicolumn{1}{|l|}{\textbf{Slope}}                   & \multicolumn{1}{r|}{0.9793}  & \multicolumn{1}{r|}{0.9884}  & \multicolumn{1}{r|}{0.9687}  & \multicolumn{1}{r|}{0.9852}  \\ \hline
\multicolumn{1}{|l|}{\textbf{Intercept}}               & \multicolumn{1}{r|}{0.0047}  & \multicolumn{1}{r|}{-0.0009} & \multicolumn{1}{r|}{0.0016}  & \multicolumn{1}{r|}{0.0027}  \\ \hline
\end{tabular}
\end{table}

\begin{figure*}[tbp]
\centering
\begin{subfigure}{0.48\textwidth}
    \centering
    \includegraphics[width=\linewidth]{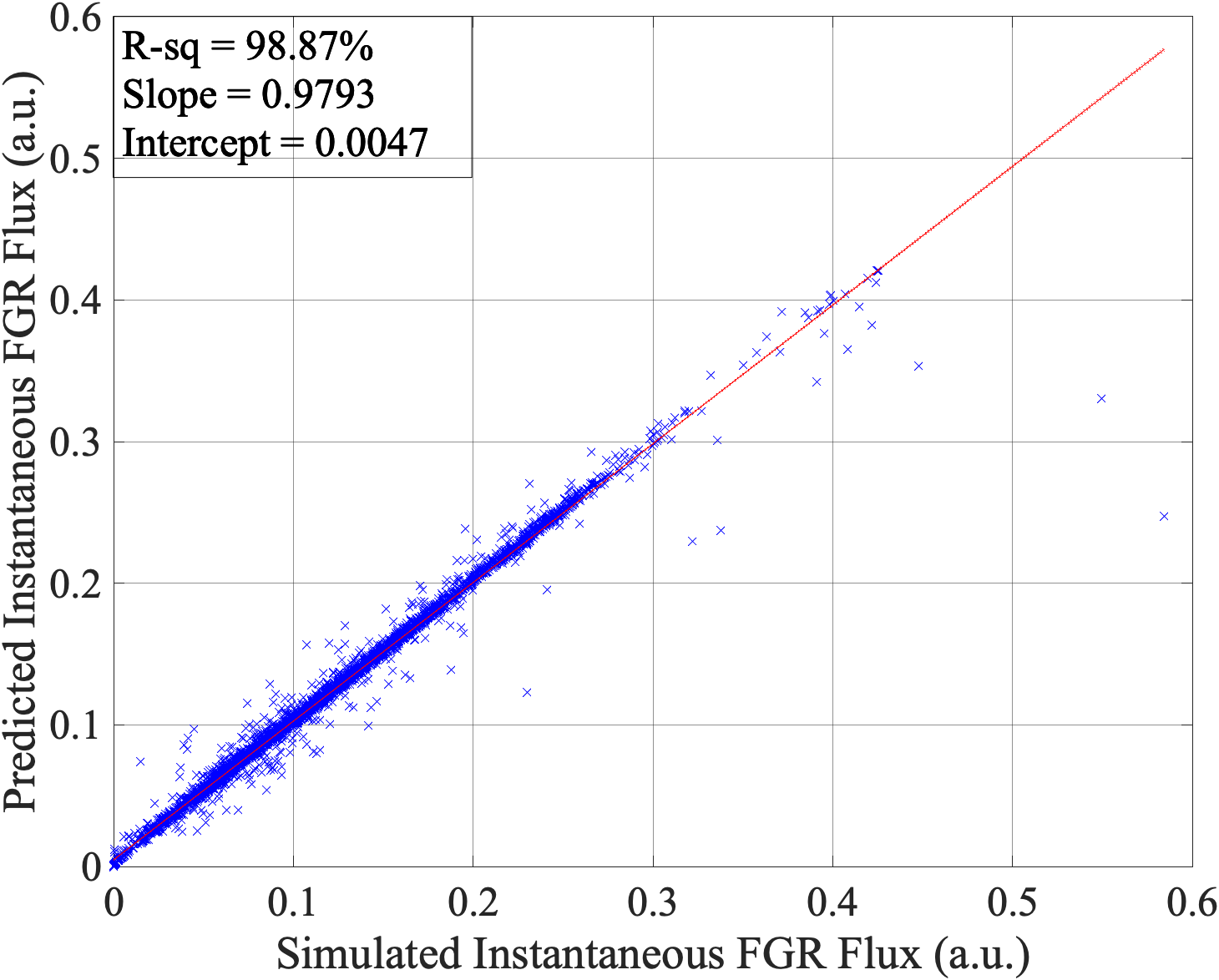}
    \caption{Baseline DenseNet}
\end{subfigure}
\begin{subfigure}{0.48\textwidth}
    \centering
    \includegraphics[width=\linewidth]{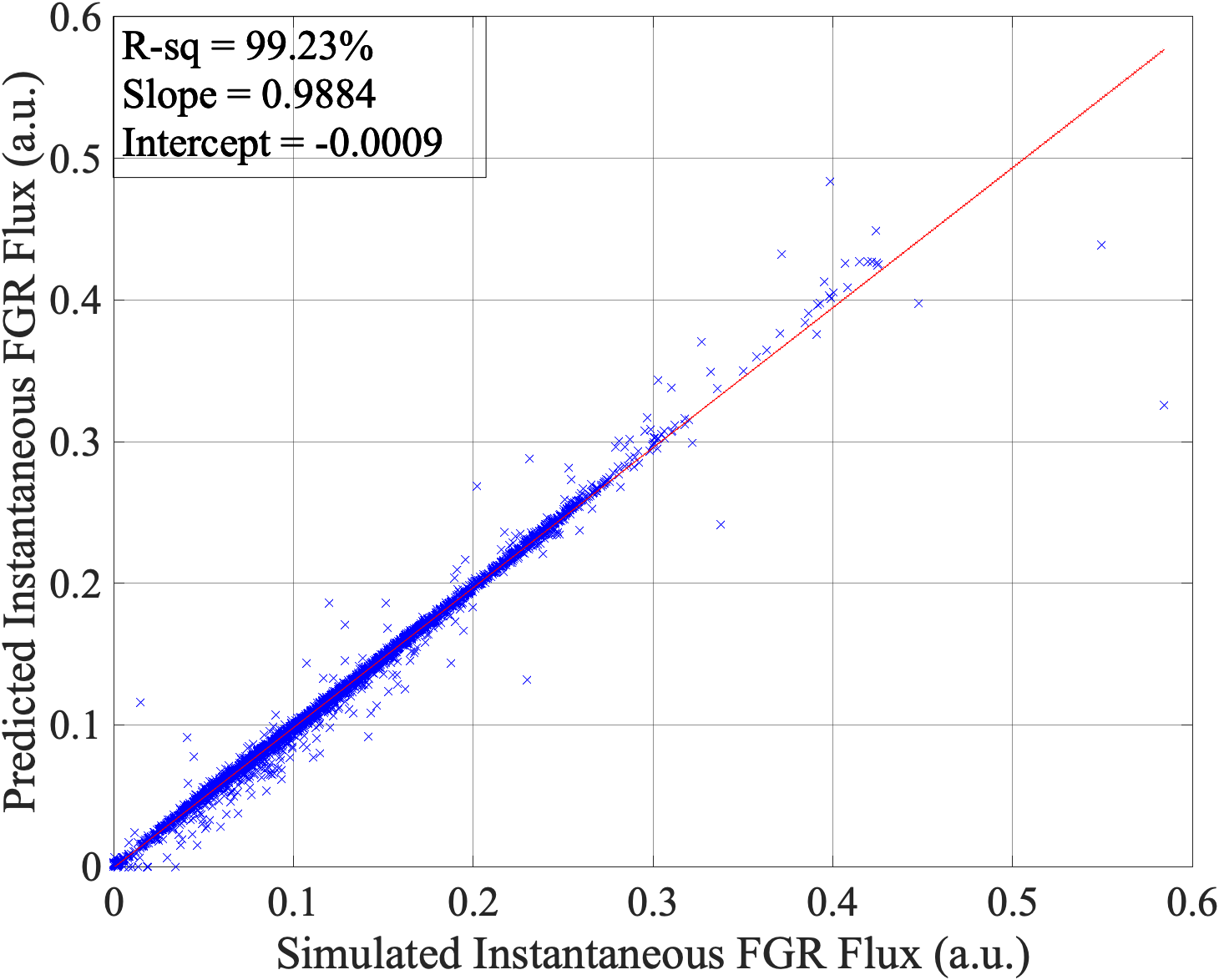}
    \caption{DenseNet + CBAM}
\end{subfigure}
\begin{subfigure}{0.48\textwidth}
    \centering
    \includegraphics[width=\linewidth]{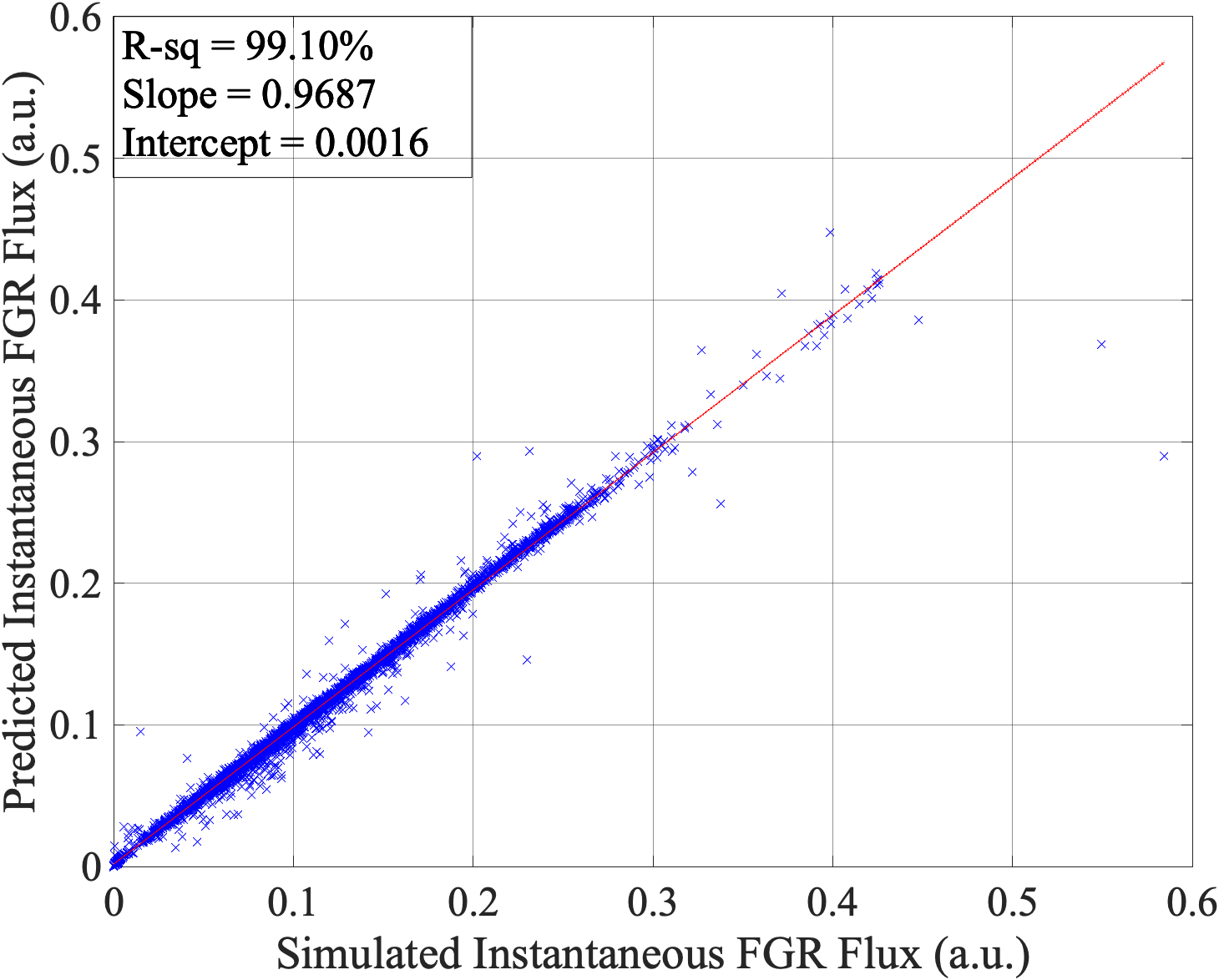}
    \caption{DenseNet Inception-A + CBAM}
\end{subfigure}
\begin{subfigure}{0.48\textwidth}
    \centering
    \includegraphics[width=\linewidth]{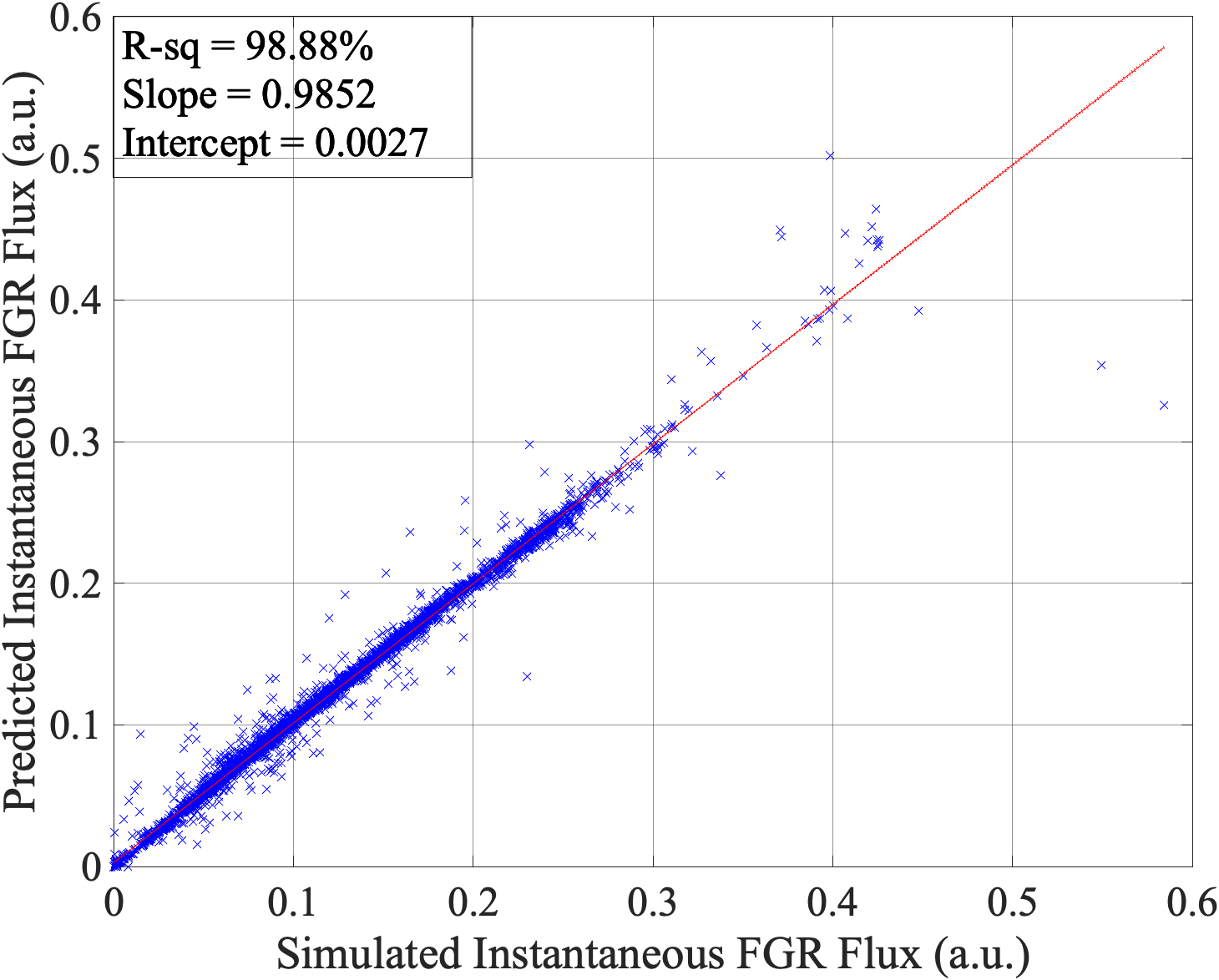}
    \caption{DenseNet Inception-B + CBAM}
\end{subfigure}
\caption{Plot of predicted vs.\ simulated instantaneous FGR flux for each examined network.}
\label{fig:9}
\end{figure*}

Figure~\ref{fig:9} illustrates the linearity of the predicted vs.\ simulated instantaneous FGR flux values for each network. All networks exhibited excellent $R^{2}$ values (above 98\%). Two networks, DenseNet + CBAM and DenseNet Inception-A + CBAM, have $R^{2}$ values above 99\%.

%\newpage

% \begin{figure*}[tbp]
%     \centering
%         \includegraphics[scale=1]{figures/InceptionA-saliency-98.png}
%     \caption{Example of Tensorflow GradientTape saliency map for Dense Inception-A + CBAM network prediction on a zero-connectivity test set image (simulated FGR flux = 0.00983, predicted FGR flux = 0.00989, absolute percentage error = 0.62\%). A blue tint on a pixel indicates greater network activation at that pixel.}
%     \label{fig:11}
% \end{figure*}

The trained neural network surrogate models can rapidly predict instantaneous FGR flux when presented with a novel microstructure image input. The following evaluations provide an indication of the computational acceleration that can be achieved comparing to the reference mesoscale multiphysics simulations. Note that the latter generate both 2D microstructure images and FGR flux values, whereas the CNN surrogate models studied generate only instantaneous FGR flux value predictions.

The mesoscale hybrid FGR model, with parameters described in section II.C, is executed on the University of Florida's HiPerGator \cite{ufrc} on an 80-CPU (Intel Xeon Gold 6142 CPU @ 2.60 GHz) partition with no GPU acceleration. The mean execution time per time step of the hybrid model is found to be approximately 26.5 seconds. The four neural network models are evaluated using two different HiPerGator configurations. The first configuration is identical to the one used in the reference multiphysics simulations, while the second configuration uses a single Intel Xeon Gold 6142 CPU @ 2.60 GHz and one NVIDIA A100 GPU. 

Table~\ref{tab:3} lists the mean execution time for instantaneous FGR flux prediction per microstructure image for the four neural network models and the two configurations. Using 80 CPUs and no GPUs, the execution time of the four neural network models varies by only around 12\%. Dense Inception-A + CBAM is the most computationally efficient. DenseNet+CBAM is the least computationally efficient. These execution times are around three orders shorter than for simulations using the hybrid model. Using 1 CPU and 1 GPU, the execution time of the four models varies by around 28\%. Baseline DenseNet and DenseNet+CBAM are the most computationally efficient for this configuration, and Dense Inception-A+CBAM is the least computationally efficient. The times with a GPU are around one order of magnitude shorter than with 80 CPUs and are four order of magnitude shorter than the hybrid model. 

\begin{table*}[btp]
\centering
\caption{Mean execution times of examined neural network architectures for predicting instantaneous FGR flux from single image inputs. Note that the execution time for one time step of the hybrid model using 80 CPUs and no GPUs is 26.5 s ($26.5\times10^3$ ms).}
\label{tab:3}
\begin{tabular}{l|llll|}
\cline{2-5}
& \multicolumn{4}{c|}{\textbf{Mean Execution Time per Image (ms)}}            \\ \hline
\multicolumn{1}{|l|}{\textbf{Configuration}} &
  \multicolumn{1}{l|}{\textbf{\begin{tabular}[c]{@{}c@{}}Baseline \\ DenseNet\end{tabular}}} &
  \multicolumn{1}{l|}{\textbf{\begin{tabular}[c]{@{}c@{}}DenseNet \\ + CBAM\end{tabular}}} &
  \multicolumn{1}{l|}{\textbf{\begin{tabular}[c]{@{}c@{}}Inception-A \\ + CBAM\end{tabular}}} &
  {\textbf{\begin{tabular}[c]{@{}c@{}} Inception-B\\ + CBAM\end{tabular}}} \\ \hline
\multicolumn{1}{|l|}{80 CPUs, No GPU} & \multicolumn{1}{l|}{25.1} & \multicolumn{1}{l|}{28.3} & \multicolumn{1}{l|}{24.7} & 26.6 \\ \hline
\multicolumn{1}{|l|}{1 CPU, 1 GPU}    & \multicolumn{1}{l|}{2.6}  & \multicolumn{1}{l|}{2.6}  & \multicolumn{1}{l|}{3.4}  & 2.9  \\ \hline
\end{tabular}
\end{table*}

% --------------------
\section{Discussion} \label{sec:discussion}
% --------------------

These results indicate that densely-connected CNNs possess a strong ability to approximate highly complex and nonlinear phenomena such as FGR in 2D microstructures. After initial training, the networks can rapidly make predictions of instantaneous FGR flux without resorting to computationally-intensive simulations using the hybrid FGR model. The network that provides overall the lowest error is Dense Inception-A + CBAM. It is also the most computationally efficient using CPUs. It is the least efficient on a GPU, but all of the times were very short. Furthermore, the studied approach is physics-agnostic and could potentially be used in other regression problems using grid-like input data and a scalar output.
\begin{figure*}[tbp]
    \centering
    \begin{subfigure}{0.43\textwidth}
        \centering
        \includegraphics[width=1\textwidth]{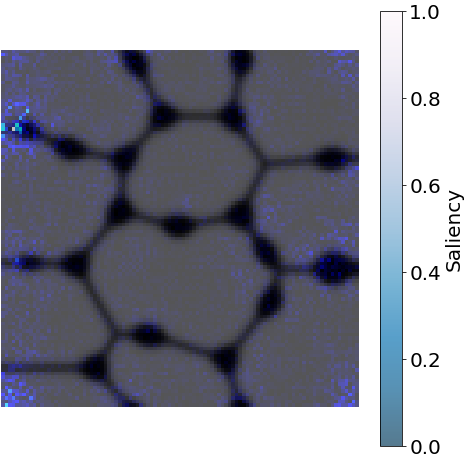}
        \caption{}\label{fig:10}
    \end{subfigure}
    \begin{subfigure}{0.43\textwidth}
        \centering
        \includegraphics[width=1\textwidth]{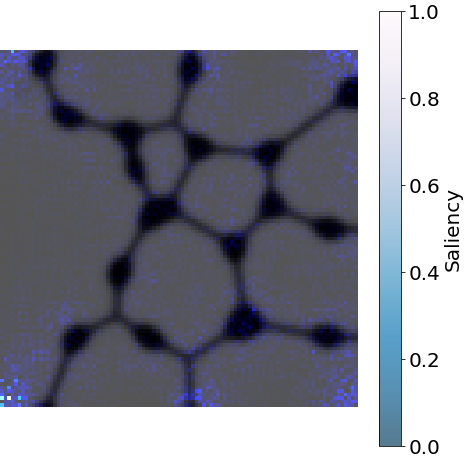}
        \caption{} \label{fig:11}
    \end{subfigure}
    \caption{Examples of Tensorflow GradientTape saliency maps for the Dense Inception-A + CBAM network prediction for (a) a case with grain boundaries contacting the free surface (simulated FGR flux = 0.13424, predicted FGR flux = 0.13456, absolute percentage error = 0.23\%), and (b) a case with no grain boundaries contacting the free surface (simulated FGR flux = 0.00983, predicted FGR flux = 0.00989, absolute percentage error = 0.62\%). The images are shaded by the saliency, which indicates the impact of that pixel on the FGR. Voids and grain boundaries are shown in black.}
\end{figure*}

Once trained, the networks predict the instantaneous FGR flux for a 2D microstructure image. In addition, the networks can provide information regarding which aspects of the microstructure have the largest impact on the FGR. This is accomplished using saliency maps \cite{Simonyan2014} generated using Tensorflow's \cite{tensorflow2015-whitepaper} GradientTape functions. Figure~\ref{fig:10} gives an example of such a saliency map. A blue tint on a pixel indicates greater impact of that pixel on the FGR. Notably, pixels on the left side of the image that are closer to the free surface have a larger impact, as do those near voids and grain boundaries. This is consistent with the fast grain boundary and surface diffusion included in the phase field model. It is expected that structures with greater grain boundary connectivity to the free surface will exhibit greater instantaneous FGR flux. Figure~\ref{fig:11} shows the saliency map for an image with a low instantaneous FGR value. The map shows that only the pixels near the bubbles on the top and bottom surfaces near the left surface have significant impact on the FGR.

The results presented in this study are subject to several limitations. The neural network models are trained on simulated data generated using only 10-grain microstructures, temperature set at 1500 K, and a free surface on the left side of the microstructure. Further work is needed to increase the network generalizability by expanding the training dataset with simulation runs carried at different temperatures and with varying free surface configurations and grain counts. The modified network architectures would need to accept temperature and free surface location as additional input parameters. Another natural extension of this work is to process 3D microstructures using 3D, rather than 2D, convolutions throughout the network layers.

This work, predicting the instantaneous fission gas release given a microstructure image, is a necessary first step in our ultimate goal of predicting the both the FGR flux and the evolution of the UO$_{2}$ microstructure over time. Such a model could be accomplished by incorporating a CNN model into a recurrent neural network. The pretrained CNN could function as the encoder in a U-Net \cite{ronneberger2015unet} architecture with Long Short-Term Memory \cite{lstm1997} layers used for the recurrent component.

% --------------------
\section{Conclusions} \label{sec:conclusions}
% --------------------

We tested four DenseNet-inspired neural network architectures modified to implement multiscale regression for predicting instantaneous FGR flux from 2D nuclear fuel microstructure images. These architectures were trained using 2D data from a hybrid phase field/cluster dynamics model of bubble evolution and FGR. We compared a baseline DenseNet configuration with an attention-enabled DenseNet and two densely-connected, attention-enabled Inception models. All four networks exhibited very high predictive power, with two configurations - DenseNet with attention and Dense Inception-A with attention - demonstrating $R^{2}$ values of above 99\%. Dense Inception-A with attention produced the lowest validation and test MAPE out of the four networks, as well as exhibiting the greatest robustness with respect to very low instantaneous FGR flux values and the greatest degree of stability during training. Saliency map visualizations showed that the trained models could also indicate which pixels have the largest impact on the instantaneous FGR. Preliminary computational time evaluations indicated the potential of surrogate neural network models to achieve several orders of magnitude acceleration comparing to mesoscale multiphysics simulations.

% --------------------
\section{CrediT Authorship Contribution Statement} \label{sec:credit}

\textbf{Peter Toma}: Investigation, Methodology, Software, Data Curation, Writing - Original Draft, Visualization. \textbf{Md Ali Muntaha}: Supervision, Conceptualization, Methodology, Formal Analysis, Writing - Review \& Editing. \textbf{Joel B. Harley}: Validation, Methodology, Writing - Review \& Editing. \textbf{Michael R. Tonks}: Project Administration, Supervision, Conceptualization, Resources, Writing - Review \& Editing

\section{Declaration of Competing Interest} \label{sec:interests}

The authors declare that they have no known competing financial interests or personal relationships
that could have appeared to influence the work reported in this paper.

\section{Acknowledgements} \label{sec:acknowledgements}

We express our gratitude for the high-performance computing resources provided by the University of Florida’s HiPerGator clusters, which facilitated the execution of computationally intensive 2D simulations.

The time of Tonks and Muntaha for this work was supported by the U. S. Department of Energy, Office of Nuclear Energy and Office of Science, Office of Advanced Scientific Computing Research through the Scientific Discovery through Advanced Computing (SciDAC) project on Simulation of Fission Gas through the grant DOE DE-SC0018359 at the University of Tennessee.

\section{Data Availability} \label{sec:availability}

The MOOSE input files used to generate the simulation results in this paper, the Python source code used to train and evaluate the networks, and the trained networks can be obtained from the authors upon reasonable request.

% --------------------

% %%%%%%%%%%%%%%%%%%%%%%%%%%%%%%%%%%%%%%%%%%%%%%%%%%%%%%%%%%
% %%%%%%%%%%%%%%%%%%%%%%%%%%%%%%%%%%%%%%%%%%%%%%%%%%%%%%%%%%
% REFERENCES SECTION
% %%%%%%%%%%%%%%%%%%%%%%%%%%%%%%%%%%%%%%%%%%%%%%%%%%%%%%%%%%
% %%%%%%%%%%%%%%%%%%%%%%%%%%%%%%%%%%%%%%%%%%%%%%%%%%%%%%%%%%
\medskip

\bibliography{references.bib} 

%\newpage

% ==========================
% ==========================
% ==========================

\end{document}